\input harvmac

\lref\bsv{R. Britto-Pacumio, A. Strominger and
A. Volovich, {\it Two Black-Hole Bound States}, JHEP 0103:050 (2001),
hep-th/0004017.}

\lref\vafgo{
R.~Gopakumar and C.~Vafa,
{\it M-theory and Topological Strings. II,}
hep-th/9812127.}

\lref\okonek{C. Okonek, M. Schneider and  H. Spindler, { \it
Vector Bundles on Complex Projective Spaces}, 1980, Boston, MA;
Basel: Birkhauser.}

\lref\hp{P. Howe and G. Papadopoulos, Comm. Math. Phys. {\bf 151}
(1993) 467.}

\lref\scqm{J. Michelson and A. Strominger, {\it The Geometry of
(Super) Conformal Quantum Mechanics,}  Commun.\ Math.\ Phys.\ {\bf 213}, 1 (2000), hep-th/9907191.}

\lref\iceland{R. Britto-Pacumio, J. Michelson, A. Strominger and
A. Volovich, {\it Lectures on Superconformal Quantum Mechanics and
Multi-Black Hole Moduli Spaces,} hep-th/9911066.}

\lref\ddf{V. de Alfaro, S. Fubini and G. Furlan, {\it Conformal
Invariance in Quantum Mechanics,} Il Nuovo Cimento {\bf 34A}
(1976) 569.}

\lref\gmt{ J.P. Gauntlett, R.C. Myers and P.K. Townsend, {\it
Black Holes of D=5 Supergravity,} Class.\ Quant.\ Grav.\ {\bf 16}
(1999) 1, hep-th/9810204.}

\lref\gt{G. W. Gibbons and P. K. Townsend, {\it Black Holes and
Calogero Models,} Phys.\ Lett.\ {\bf B454} (1999) 187,
hep-th/9812034.}

\lref\cdkk{P. Claus, M. Derix, R. Kallosh, J. Kumar, P.K. Townsend
and A. Van Proeyen, {\it Black Holes and Superconformal
Mechanics,} Phys.\ Rev.\ Lett. {\bf 81} (1998) 4553,
hep-th/9804177.}

\lref\jmas{J. Michelson and A. Strominger, {\it Superconformal
Multi-Black Hole Quantum Mechanics,} JHEP  9909:005 (1999),
hep-th/9908044.}

\lref\gpjg{J. Gutowski and G. Papadopoulos, {\it The Dynamics of
Very Special Black Holes,} Phys.\ Lett.\ {\bf B472}, 45 (2000), hep-th/9910022.}

\lref\hull{C. M. Hull, {\it The Geometry of Supersymmetric Quantum
Mechanics,} hep-th/9910028.}

\lref\myertown{
J.~P.~Gauntlett, R.~C.~Myers and P.~K.~Townsend,
{\it Black Holes of D = 5 Supergravity,}
Class.\ Quant.\ Grav.\  {\bf 16}, 1 (1999),
hep-th/9810204.
}

\lref\pap{G. Papadopoulos, {\it Conformal and Superconformal
Mechanics,} hep-th/0002007.}

\lref\mss{A. Maloney, M. Spradlin and A. Strominger, {\it
Superconformal Multi-Black Hole Moduli Spaces in Four Dimensions,}
hep-th/9911001.}

\lref\gutpap{J. Gutowski and G. Papadopoulos, {\it Moduli Spaces
for Four-and Five-Dimensional Black Holes,} Phys.\ Rev.\ {\bf D62}, 064023 (2000), hep-th/0002242.}

\lref\mms{J. Maldacena, J. Michelson and A. Strominger, {\it
Anti-de Sitter Fragmentation,} JHEP 9902:011 (1999),
hep-th/9812073.}

\lref\iceland{R. Britto-Pacumio, J. Michelson, A. Strominger and
A. Volovich, {\it Lectures on Superconformal Quantum Mechanics and
Multi-Black Hole Moduli Spaces,}
hep-th/9911066.}

\lref\asbhs{
A.~Strominger,
{\it Black Hole Statistics,}
Phys.\ Rev.\ Lett.\  {\bf 71}, 3397 (1993),
hep-th/9307059.
}

\lref\amspvt{A. Maloney and M. Spradlin, private communication.}

\lref\atiyah{
M.~F.~Atiyah and N.~J.~Hitchin,
{\it Low-Energy Scattering Of Nonabelian Monopoles,}
Phys.\ Lett.\ A {\bf 107}, 21 (1985).
}
\lref\manton{
G.~W.~Gibbons and N.~S.~Manton,
{\it Classical And Quantum Dynamics Of BPS Monopoles,}
Nucl.\ Phys.\ B {\bf 274}, 183 (1986).
}

\lref\sen{A.~Sen,
{\it Dyon - Monopole Bound States, Selfdual Harmonic Forms on the Multi - Monopole Moduli Space, and SL(2,Z) Invariance in String Theory,}
Phys.\ Lett.\ B {\bf 329}, 217 (1994),
hep-th/9402032.
}

\lref\ferr{
R.~C.~Ferrell and D.~M.~Eardley,
{\it Slow Motion Scattering And Coalescence Of Maximally Charged Black Holes,}
Phys.\ Rev.\ Lett.\  {\bf 59}, 1617 (1987).
}

\lref\gibb{
G.~W.~Gibbons and P.~J.~Ruback,
{\it The Motion Of Extreme Reissner-Nordstrom Black Holes In The Low Velocity Limit,}
Phys.\ Rev.\ Lett.\  {\bf 57}, 1492 (1986).
}

\lref\jm{
J.~Maldacena,
{\it The Large N Limit of Superconformal Field Theories and Supergravity,}
Adv.\ Theor.\ Math.\ Phys.\  {\bf 2}, 231 (1998), hep-th/9711200.
}

\lref\krantz{
S.~Krantz,
{\it Function Theory of Several Complex Variables}, 
John Wiley \& Sons, 1982. 
}

\def\p{\partial}
\def\bp{{\bar{\partial}}}

\def\bz{{\bar{z}}}
\def\ba{{\bar{a}}}

\def\cl{{\cal L}}

\def\lra{\longrightarrow}
\def\ra#1{{\buildrel{#1}\over\longrightarrow}}
\def \half{{1 \over 2}}
\def\pd{{\p^{\dagger}}}
\def\m{{\cal M}}
\def\x{{\vec x}}


\Title{\vbox{\baselineskip12pt\hbox{hep-th/0106099}\hbox{HUTP-01/A032}\hbox{}}
}{Spinning Bound States of Two and Three Black Holes}

\centerline{Ruth Britto-Pacumio$^{*,\sharp}$, 
Alexander Maloney$^*$,
Mark Stern$^\dagger$ 
and Andrew Strominger$^*$
}
\bigskip\centerline{$^*$Department of Physics}
\centerline{Harvard University}
\centerline{Cambridge, MA 02138}

\bigskip\centerline{$^\sharp$Institute for Theoretical Physics}
\centerline{University of California}
\centerline{Santa Barbara, CA 93106}

\bigskip\centerline{$^\dagger$Department of Mathematics}
\centerline{Duke University}
\centerline{Durham, NC 27706}

\vskip .3in \centerline{\bf Abstract} {Bound states of BPS
particles in five-dimensional ${\cal N}=2$ supergravity are
counted by a topological index.  We compute this bound state index
exactly for two and three black holes as a
function of the $SU(2)_L$ angular momentum. The required regulator
for the infrared continuum of near-coincident black holes is
chosen in accord with the enhanced superconformal symmetry.}

\smallskip
\Date{}
\listtoc
\writetoc

\newsec{Introduction}

The study of semiclassical soliton scattering and moduli spaces
has a long and rich history. A beautiful chapter, relevant to the
present work, began with the realization that a pair of
slowly-moving supersymmetric BPS monopoles is described by quantum
mechanics on the two-monopole moduli space, which turns out to be
the Atiyah-Hitchin space \refs{\manton, \atiyah}.  The number of
bound states is then determined by the moduli space cohomology,
and is in agreement with predictions from S-duality \sen.

It is natural to try to develop a similar picture for
supersymmetric black holes. This problem is especially interesting
because it provides a new angle to study the deep puzzles
associated to quantum mechanical black holes. Work on construction
of the $N$-black hole moduli space, which we shall denote
$\m_N$,  began in the early eighties
\refs{\gibb, \ferr}. However the black hole problem turns out to
be considerably more subtle than its monopole counterpart, in
part because of divergences near the horizon at intermediate
stages of the calculation. The supersymmetric moduli space for $N
\ge 3$ has been found only very recently \refs{\jmas \gpjg \mss
-\gutpap}.\foot{The moduli space in \ferr\ is inconsistent with
supersymmetry at sixth order in the black hole masses 
and was corrected in \mss.}

Now that the moduli space is known, it is natural to try to
compute the number of bound states of $N$ black holes or, more
reliably, the supersymmetric bound state index
\eqn\csfg{{\cal
I}^{(N)}(y)={\rm Tr}~y^{2J^3_L}(-)^{2J^3_R},}
where the trace is over all states in the $N$ black hole quantum mechanics
and  $(J^3_L, J^3_R)$ are $SU(2)_L\times SU(2)_R$ angular 
momentum operators. Here one
immediately encounters a puzzle.  The moduli space quantum
mechanics contains a divergent continuum of states describing
highly redshifted, near-coincident black holes.\foot{Near-horizon
infrared divergences of this type have appeared in a variety of
contexts in black hole physics. The presence of a
divergent continuum of states, with a  naively infinite
capacity for information storage, is
closely related to the information puzzle. Therefore we
expect a proper understanding of how to regulate this infrared
divergence to be relevant to the information puzzle.} In order to
compute ${\cal I}^{(N)}$ one must regulate this continuum. It is not
obvious how the regulator should be chosen.

A relevant discovery in  \jmas\ is that this infrared continuum
of states is in a representation of an enhanced
superconformal symmetry. This observation is obviously pertinent
to an understanding of the still-enigmatic $AdS_2/CFT_1$
correspondence, but the precise connection remains
mysterious.\foot{The
enhanced superconformal symmetry group of the black hole quantum mechanics is
the same as the superisometry group of $AdS_2\times S^3$, indicating some
relation of the former to 
the sought $CFT_1$ of $AdS_2/CFT_1$. However the $CFT_1$ envisioned in
\jm\ was e.g.\ the quantum mechanics of a wrapped D-brane moduli space of
the type considered in \vafgo, which on
the face of it is rather different.}
In \bsv\ it was shown that the superconformal symmetry singles out a
natural regulator for the infrared continuum. This regulator was
then used to relate the index to `superconformal cohomology' on the
moduli space. Superconformal cohomology employs the nilpotent operator
$\p-D$, where the $(1,0)$ form $D$ is associated to conformal scale
transformations on $\m_N$. The
cohomology was computed in \bsv\ for the simplest case of two
supersymmetric black holes in five dimensions with internal
$SU(2)_L\times SU(2)_R$ spin $(0,\half)$.

In the present paper we present a solution of the
superconformal bound state problem for two and three BPS black holes in five
dimensions as a function of the spin eigenvalues
$(j_L,j_R)$.  It is shown that for the $N$ black hole problem 
the superconformal cohomology vanishes except at rank
$N-1$. A cohomology generating function is then
defined by \eqn\cfg{{\cal
Z}^{(N)}(y,z)=\sum_{j_L,j_R} {\rm dim}H^{N-1}(\m_N,j_L,j_R)  
y^{2j_L}z^{2j_R},
} 
where the sum is over all
the superconformal chiral primary states in the $N$-black hole
quantum mechanics.  We find 
\eqn\cdfg{\eqalign{
{\cal Z}^{(2)}(y,z) &= {yz \over (1-yz)(y-z)} \cr
{\cal Z}^{(3)}(y,z) &= 2 \left[{yz \over (1-yz)(y-z)}\right]^2 .
}}
The generating function ${\cal Z}^{(N)}$ is not (as far as we know)
invariant under supersymmetric corrections to the black
hole quantum mechanics, so it is more natural to 
use the index ${\cal I}^{(N)}$ found by setting $z=-1$ in \cdfg.
This index is invariant under all deformations that preserve the superconformal
structure of the quantum mechanics.

This paper is organized as follows. Section 2 contains a review of
the geometry of the black hole moduli space, the superconformal structure,
and the relation of the index to moduli space cohomology.
This material is largely from \refs{\jmas, \bsv} and much of it is reviewed
in \iceland.
In section 3 we show that the cohomology lives only in the middle
dimension.  In section 4 we calculate the bound state indexes
${\cal I}^{(2)}$ and ${\cal Z}^{(2)}$
for two black holes, generalizing the result of \bsv.
In section 5 we prove two more vanishing theorems that reduce the problem
to the consideration of cohomology classes on certain subsets of the moduli
space.  In the three black hole case this cohomology may be found 
exactly using several Mayer-Vietoris type arguments, allowing us to
compute ${\cal I}^{(3)}$ and ${\cal Z}^{(3)}$. 
In section 6 we raise the issue of uniqueness of our 
adopted definition of the index. We discuss an alternate definition which 
does not involve 
the superconformal structure and gives a trivial result (at least for 
$N=2$).

\newsec{Review of Superconformal Black Hole Quantum Mechanics}

In this section we review pertinent results on the quantum
mechanics of slowly moving black holes. We follow the notation of
\bsv\ where many of the statements are derived in more detail.

Consider $N$ slowly moving BPS black holes in five-dimensional
${\cal N}=2$ supergravity with no matter. For
sufficiently slow motion, the dynamics governing the relative
positions is described by a quantum mechanical sigma model whose
target space is the $N$-black hole moduli space.\foot{For
simplicity we here and hereafter ignore the center-of-mass degrees
of freedom.}
 Points on $\m_N$ are parameterized by the
relative positions of the $N$ black holes.\foot{
In the $N$-monopole problem, the moduli space has an asymptotic
identification under the permutation group $S_N$, corresponding to
the fact that the monopoles are identical particles. This
identification is required for smoothness of the moduli space in
the interior. In the black hole case, in contrast, the moduli
space is smooth without identifications.  Implementing $S_N$
identifications could induce extra cohomology above and beyond
what we find herein.
Whether or not this is appropriate may depend on
microscopic considerations and cannot be semiclassically
determined \asbhs. In this paper we do not consider such
identifications.}
The quantum mechanics has four linearly realized Poincare
supersymmetries inherited from the four spacetime symmetries which
are unbroken by the BPS black holes. It also has an
$SU(2)_{L} \times SU(2)_{R}$ global symmetry
arising from spatial rotations and an additional $R$-symmetry which we denote 
$SU(2)_I$.\foot{In \bsv\ this $R$-symmetry was denoted $SU(2)_R$.}

At very low energies one finds that the theory splits into two
different types of  decoupled sectors.  One describes
noninteracting,  freely moving black holes while the other
describes strongly interacting, near-coincident black holes. The
near-coincident quantum mechanics has an enhanced superconformal
symmetry $D(2,1;0)$ which has eight supercharges and incorporates
$SU(2)_{R}$ spatial rotations \refs{\myertown,\jmas}. We denote the near-coincident
moduli space $\m_N$. The superconformal structure highly
constrains the geometry of $\m_N$ \scqm\ as we now describe.

\subsec{The Geometry of $\m_N$}

The near-coincident $N$-black hole moduli space $\m_N$ has a
triplet of self-dual complex structures obeying
\eqn\rtyo{I^rI^s=-\delta^{rs}+\epsilon^{rst}I^t,} for $r,s=1,2,3$.
The complex dimension is given by \eqn\dimm{ {\rm dim_C
}(\m_N)=2N-2=n.} The metric is \eqn\met{g_{a \bar
b}=\half\bigl(\p_a\p_{\bar b}L + I^{-\bar c}_aI^{+d}_{\bar
b}\p_{\bar c}\p_d L).} In this expression, the complex coordinates
\eqn\dcml{z^a, z^{\ba},~~~~~~~~~~a=1,2,...,n} are adapted to
$I^3$, and $I^\pm=\half (I^1\pm i I^2)$. The $z^a$ are built out of
the real coordinates of the black holes $\vec X^A, A=1,...,N$
after factoring out the center of mass in the usual way. We will
use indices $a,b,...$ for the $n$ complex coordinates and
$M,N,...$ for the $2n$ real coordinates. $L$ is a function of the
black hole positions given by\foot{For simplicity we have set all the
black hole charges equal to $Q$.} \eqn\gtf{L=-\int d^4 X \bigl(
\sum_{A=1}^N {Q \over |\vec X- \vec X^A|^2}\bigr)^3.} Although
this function is at first sight divergent, the infinite part of
$L$ does not contribute to the metric \met---we refer the reader
to appendix A for details.  Removing this
irrelevant part, we find that $L$ obeys \eqn\lpa{(z^a\p_a + 1)L = -\half
K} where $K$ is the function \eqn\kfun{K=6\pi^2 \sum^N_{A\ne
B}{Q^3\over |\vec X^A - \vec X^B|^2}.} The metric \met\ has a
complex homothety generated by the Lie derivative
${\cal L}_D g_{a {\bar b}} = g_{a {\bar b}}$,
which acts on scalars as \eqn\hmth{{\cal
L}_D \sim -z^a\p_a . } We use $D^a$
($D^{\bar a}$) to denote the homothetic vector field $-z^a$
($-z^{\bar a}$) and $\bar D$ ($D$) to denote the associated
$(0,1)$ ($(1,0)$) form constructed with the metric $g_{a \bar b}$.
The norm of $D$ is just \eqn\nmd{D_aD^a=K.}

The imaginary part of \hmth\ is part
of an $SU(2)_{R} $ triplet of isometries generated by
\eqn\rtgi{{\cal L}_{D^r}\sim X^MI^{rN}_M\p_N,~~~~~~ r=1,2,3 .}
There are
also $SU(2)_{L}$ isometries generated by
\eqn\rtyh{X^MK^{rN}_M\p_N,~~~~~~ r=1,2,3, }
where $K^r$ are the triplet of
anti-self dual complex structures.

Using the metric and the complex structures, one may construct
holomorphic and antiholomorphic two-forms
\eqn\zmxf{I^-=\half I^-_{ab}dz^adz^b,~~~~~~I^+=\half I^+_{\bar a \bar
b}dz^{\bar a}dz^{\bar b}.}  $I^-$ obeys the relations \eqn\daz{\p I^-=0,
~~~~
\p
* e^{-\phi} I^-=0,} with $*$ the Hodge dual.\foot{In contrast to
 the conventions of
\bsv, we here include complex
conjugation.}  In the $\vec X^A$ coordinates $\phi$ is given by
\eqn\ftlz{e^{2\phi}=\sqrt{
{\rm det}g}.}

\subsec{The Hilbert Space as $(p,0)$-Forms}

The Hilbert space of the black hole quantum mechanics can be
identified with the space of $(p,0)$-forms $f_p$ on $\m_N$. The
inner product is \eqn\inp{\langle f^\prime_p
|f_p \rangle =\int_{\m_N}e^{-\phi-2K}{f}^\prime_p \wedge *f_p.} The action
of the superconformal algebra is simply represented on $(p,0)$-forms. $D(2,1;0)$ is the semidirect product of $SU(1,1|2)$ and
$SU(2)_I$. The $SU(2)_I$ is generated by the Lefschetz
action of ${ i \over 2}I^-$,
\eqn\vcx{\eqalign{
	{ i \over 2}I^3 f_p&={n -2p\over 4} f_p,\cr
{ i \over 2}I^-f_p&={ i \over 2 } I^-\wedge f_p,\cr
{ i \over 2}I^+f_p&=-{ i \over 2} *I^-\wedge * f_p,}}
 The bosonic $R$-symmetry in
$SU(1,1|2)$ is $SU(2)_{R}$, and is generated by the Lie
derivatives \rtgi. The supercharges are in the $(2,2)$ of
$SU(2)_{R}\times SU(2)_I $. The operator $L_0$ is the
hamiltonian associated to the metric \met\ with potential \nmd.
The actions of the weight $(\half, -\half)$ charge and its adjoint
with respect to \inp\ are \eqn\vcxp{\eqalign{G^{+-}_{- \half}f_p&=-{i\over \sqrt{2}}(\p
f_p-2D\wedge f_p) ,\cr G^{-+}_{\half}f_p&={i\over \sqrt 2} \pd f_p
.}} where
\eqn\zmv{\pd \equiv -*e^\phi  \p e^{-\phi}*} Commutators of
these basic operators  then generate the full algebra.

\subsec{Chiral Primaries }

A chiral primary is a normalizable $(p,0)$-form
that is annihilated by both $G^{+-}_{-\half}$ and its adjoint
$G^{-+}_{\half}$. From \vcxp\ this is equivalent to the harmonic condition
\eqn\gjjs{\p f_p-2D\wedge f_p = \pd f_p=0.}
Thus the chiral primaries are harmonic representatives of $H^p(\m^N)$,
which we define to be the cohomology of normalizable
$(p,0)$-forms on $\m^N$ relative to the differential operator $\p-2D$.
The $D(2,1;0)$ algebra can be used to show that
a normalizable solution of \gjjs\ is annihilated by $J_R^+$,
$L_0-J^3_R$, $L_1$ and all the supercharges except for
$G_{-\half}^{-\pm}$. This implies the relations \eqn\nwv{\p f_p
=0,} \eqn\czm{D\wedge f_p=0,}
\eqn\dzkl{ (D^{
a}\p_{a}- D^{\bar a}\p_{\bar a})f_p=2j_Rf_p,}
\eqn\zkl{ D^{\bar a} I^{+b}_{\bar a}
\p_{b}f_p=0,} where $j_R$ is the eigenvalue of and $f_p$ is an eigenform of
$J^3_R$. 
These last two equations follow from our choice of complex coordinates
adapted to $J^3_R$ in which $D^a=-z^a$.
Further useful relations may be found in \bsv. A chiral
primary with $j_R=0$ would be an $SU(2)_{R}$ singlet
annihilated by all eight supercharges. It is easy to see that this
is impossible. Hence $j_R$ is strictly positive.

\subsec{The Bound State Index}

For theories with eight supercharges in five dimensions, the
weighted numbers of
BPS states are given by an index---roughly the difference in the numbers of
hypermultiplets and vector multiplets. This index is invariant under smooth
deformations of the theory. It has been computed in some examples
of
M-theory compactifications in \vafgo.
For the case
of multi-black hole bound states, a prescription must be given for
regulating the infrared continuum of near-coincident black holes.
In \bsv\ a regulator was proposed which amounts to
working in a basis of $L_0$ rather than Hamiltonian eigenstates.
$L_0$ differs from the Hamiltonian by the potential $K$ (equation \kfun)
which eliminates the infrared divergences.
With this prescription computing the
index reduces
to
counting the chiral primaries, weighted by $(-)^{2j_R}$.
 Since
the $SU(2)_L$ generators $J_L$ commute with all the superconformal
generators, the index can be refined by
inclusion of a $J^3_L$ weighting factor
\eqn\indsx{{\cal I}^{(N)}(y)={\rm Tr}(-)^{2J^3_R}y^{2J^3_L}.}
The quantum mechanics on $\m_N$ and hence
expression \indsx\ does not include
the center of mass degrees of freedom of the $N$ black holes.
Including this would multiply \indsx\ by a factor of
$(y^\half +y^{-\half})^2$.

\newsec{A Vanishing Theorem}

In this section, we show that chiral primaries $f_{p}$ exist only
at middle rank.  That is, $p={n \over 2}=N-1$.   We will take $f_p$ to
be a $J^3_R$ eigenform with eigenvalue $j_R$ and hence obeying
\dzkl. Our strategy will be to first  show that all chiral
primaries are annihilated by $I^-$.

Given a chiral primary $f_{p}$, it is straightforward to verify
that the action of $I^-$ \eqn\fanl{I^-\wedge f_{p}} generates a
new chiral primary (if it does not vanish). Using $I^-=- \p \p^- L$,
with $\p^- \sim dz^aI^{-\bar b}_a \p_{\bar b}$ and $L$ the finite
expression in appendix A,
the norm is
 \eqn\iwnrm{\langle I^-\wedge f_{p}|I^-\wedge f_{p}
\rangle =\int e^{-\phi-2K} ( \p \p^-L\wedge  f_{p})
\wedge * \p \p^-L\wedge f_{p} .} If boundary terms can be ignored,
this integral vanishes upon integration by parts with respect to
$\p$.  Note that the factor of $e^{-2K}$ in the integral insures that the
boundary terms vanish at large $K$.

We wish to show that the boundary terms vanish at small $K$ as well. 
To do this, we will
introduce cutoff functions into the integral and then show that the
error terms they introduce can be made arbitrarily small.
Let
$\rho_\mu :(0,\infty)\rightarrow [0,1]$ be a sequence of
differentiable compactly supported cutoff functions
such that $\rho_\mu(t) = 1$ for $t\in [{1 \over \mu}, \mu]$
and such that
\eqn\tlt{ |\rho_\mu'(t)| < {1 \over t|\ln t|}.}
With these hypotheses, we have
\eqn\iwnrp{\langle I^-\wedge f_{p}|I^-\wedge f_{p}\rangle =
\lim_{\mu \rightarrow\infty}\langle \p\p^-L\wedge f_{p}|\rho_\mu(K)I^-\wedge
f_{p}\rangle.}
The cutoff function allows us to integrate by parts obtaining
\eqn\iwnrq{\lim_{\mu \rightarrow\infty}\langle \p^-L\wedge
f_{p}|[\p^\dagger ,\rho_\mu(K)]I^-\wedge f_{p}\rangle.}
The derivative of the cutoff function is converging to zero pointwise.
Hence, (given the finiteness of $||f_p||^2$)
 the above limit vanishes if we can show that
$|\p^-L||[\p^\dagger ,\rho_\mu (K)]|$ is bounded independent of $\mu$.

First we note using \nmd\ that
\eqn\iwnrs{|[\p^\dagger,\rho_\mu (K)]| = |\rho_\mu'(K)\p K| < {1 \over K^{1/2}
|\ln K|}.}
On the other hand,
we show in appendix A that
\eqn\eest{{|\p^-L|\over K^{1/2}|\ln K|} < c_N,}
for some constant $c_N$ depending only on the number of black holes.

Thus,  the integral \iwnrm\ vanishes. Since the
integrand is nonnegative it must vanish pointwise. Consistency
then demands that all chiral primaries obey \eqn\psx{I^-\wedge
f_{p} =0. } Comparing with \vcx\ we see that this is equivalent to
the statement that $f_{p}$ is a lowest weight $SU(2)_I$ state
annihilated by $I^-$. On the other hand, since $SU(2)_I$ mixes the
supercharges which annihilate a chiral primary only among
themselves, chiral primaries are representations of  $SU(2)_I$. It
then follows from \psx\ that chiral primaries must be
$I$-singlets, and hence have \eqn\mjk{p={n \over 2}.}This can be
expressed as the vanishing theorem for superconformal cohomology
\eqn\vns{H^p(\m_N)=0,~~~~p \neq
N-1.}

\newsec{The Index for Two Black Holes}

For $N=2$ we can find the chiral primaries
(i.e.\ the group $H^1(\m_2)$) by direct computation.
The metric is
\eqn\nmel{
ds^2=2 c
{d\vec X \cdot  d \vec X \over |\vec X|^4}}
where $\vec X \equiv \vec X^1-\vec X^2$ is the relative separation
of the black holes, and $c= 12 \pi^2Q^3$ is a constant.
We have
\eqn\dft{ \vec D = - \vec X,~~~~~~
K=c |\vec X|^{-2}, ~~~~~
e^{-\phi}=c^{-1} |\vec X|^4   .}
We denote the usual complex coordinates built out of the $\vec X$ as
$z^1$ and $z^2$.

For $n=2$, we find that the chiral primary conditions \nwv\--\zkl\ are satisfied only if $f= \alpha K^{2j_R-1} D$ for some
function $\alpha$ obeying \eqn\alp{\p \alpha = 0 , ~~~~~ D^\ba
\p_\ba \alpha = (1-2j_R)\alpha.} We can calculate the norm \inp\
\eqn\redf{\eqalign{ \langle f|f\rangle &=  \int d^{4}X \sqrt g
e^{-\phi} |\alpha K^{2j_R-1}|^2 K e^{-2K} \cr &= c^{4j_R} \int d^4X
|\alpha|^2 |\vec X  |^{-2-8j_R} e^{- 2 c |\vec X |^{-2}} \cr &= c^{4j_R} \int dr
r^{-1-4j_R} e^{- 2 c r^{-2}} \int_{S^3} d^3\Omega |{\hat \alpha}
|^2 \cr &=  \half({c \over 2})^{2j_R} \Gamma(2 j_R) \int_{S^3} d^3\Omega
|{\hat \alpha}|^2 }} where $\hat \alpha$ denotes $\alpha$
restricted to the unit 3-sphere $S^3 = \{|\vec X|=1\}$. Thus $j_R \ge
\half$ and $\hat \alpha$ must be integrable on $S^3$. Clearly
$\alpha$ must be a homogeneous polynomial in $\bz^1$ and $\bz^2$
of order $2j_R-1$, since singularities of the form $(\bz^1)^{-1}$
and $(\bz^2)^{-1}$ are irregular and would cause the norm to diverge.  
There is a
basis of $2j_R$ such polynomials: $ \alpha = (\bz^1)^{2j_R-1},
(\bz^1)^{2j_R-2} (\bz^2),...,(\bz^2)^{2j_R-1}$. We may choose our basis 
of $J_L$ generators so that these polynomials have $J^3_L$ eigenvalues
$j_R-{1 \over 2}, j_R-{3 \over 2},...,-j_R+{1 \over 2}$, respectively. 
So at level $j_R$
there are $2 j_R$ chiral primaries that form an irreducible
$SU(2)_L$ multiplet of maximal spin $j_L = j_R-{1\over 2}$.
This is
summarized by
 \eqn\trg{\eqalign{{\rm dim}H^{1}(\m_2,j_L, j_R)&=1,~~~j_R \geq \half,~~~|j_L|
<j_R, ~~~ {\rm and} ~ j_L+j_R+\half\in{\bf Z} \cr
                                     &=0~~~{\rm otherwise}.}}
We adopt here the notation that $H^p(\m_N,j_L,j_R)$ is the
restriction of the cohomology  to $J^3_L$ and $J^3_R$
eigenspaces.
This reproduces the result of \bsv\ for
$j_R=\half$. Summing the chiral primaries with fixed $j_R$ weighted by
$y^{2J^3_L}$ gives
\eqn\rtg{(-)^{2j_R}\sum_{k=-j_R+\half}^{k=j_R-\half}y^{2k}=(-)^{2j_R}{y^{{2j_R}}-
y^{-2j_R}
\over y-y^{-1}}.} The index \indsx\ for $N=2$ is then obtained by
summing over $j_R$: \eqn\indf{{\cal I}^{(2)}(y)={\rm
Tr}(-)^{2J^3_R}y^{2J^3_L} =-{1 \over (y^\half+y^{-\half})^2}.}

The index \indf\ does not generate all of the cohomology because of
the unweighted sum over $j_R$. A generating partition
function for all the cohomology
can be defined by
\eqn\isdf{{\cal Z}^{(2)}(y,z)=\sum_{j_L,j_R}
{\rm dim}H^{1}(\m_2,j_L, j_R)z^{2j_R}y^{2j_L}.}
This partition function is not in general a
supersymmetric invariant index of the black hole
quantum mechanics, but nevertheless usefully summarizes
the results of our computation. For $N=2$ we have
\eqn\ikdf{{\cal Z}^{(2)}(y,z)={yz \over (1-yz)(y-z)}.}

\newsec{The Index for Three Black Holes}

The computation of the index for $N>2$ black holes is considerably more
involved. 
We first prove two more vanishing theorems that hold in the general 
$N$ case.  We then apply these to the case of three black holes and,
using several Mayer-Vietoris type arguments, compute the bound state index.

\subsec{Two More Vanishing Theorems}

  In this subsection we consider
two appropriately defined regions $V_N$ and $W_N$ of $\m_N$ 
and find that the Neumann cohomology on these subsets is trivial
for positive $j_R$.
Here, $V_N$ is the region near the
singularities of the function $K$ (i.e.\ near-coincident black
holes) and $W_N$ is the region where $K$ is small (i.e.\ widely separated black holes). 
This result will enter into
the exact sequence for the cohomology derived in the next
subsection.

It is convenient to work in terms of rescaled forms
\eqn\yti{h_p=e^{-K}f_p,} which we take to obey Neumann
boundary conditions on the region
$V_N$, so that the pullback of $*h_p$ to the boundary of $V_N$
vanishes.\foot{This condition on $*h_p$ follows from the requirement that 
$h_p$ be in the domain of $\p^\dagger$, i.e.\ for all $p-1$ forms $g$ 
on $V_N$ we have $\langle\p g, h\rangle  < c |g|$ for some ($g$ independent) $c$.
}  
The inner product \inp\ reduces to
\eqn\insp{\langle h^\prime
|h \rangle_{V_N}=\int_{V_N}e^{-\phi}h^\prime_p \wedge *{h}_p.}
If $f_p$ is a cohomology element relative to $\p-2D$ then
$h_p$ is a cohomology element relative to $\p-D$, i.e.
\eqn\hob{ (\p -D) h_p=0.}
We consider the functional $E(h_p)$ defined as
\eqn\cqd{\eqalign{
E&=||(\p -D) h_p||_{V_N}^2+||(\p^\dagger-i_D) h_p||_{V_N}^2 \cr
                    &=||\p h_p||_{V_N}^2+||\p^\dagger h_p||_{V_N}^2 +
\langle h_p \{ D, i_D\} | h_p \rangle_{V_N}-2{\rm Re} \langle h_p \{ \p, i_D\} | h_p \rangle_{V_N} \cr}}
where $i_D = *D*$ is the adjoint of the wedge product with
$D=\p K$. 
In this expression the norm is determined from \insp. 
In the second line a boundary
term which vanishes due to the Neumann condition has been
dropped in integrating by parts. Using
$ \{ \p ,
i_D\}= {\cal L}_D $, $\nabla_a e^{-\phi}D^a=0$, and integrating by
parts the last term in \cqd\ can be written 
\eqn\rwrt{ -2 {\rm Re}\langle h_p
|{\cal L}_Dh_p\rangle_{V_N}= -2j_R||h_p||^2_{V_N}-{\rm Re} \int_{\p
V_N}*De^{-\phi}|h_p|^2.} In writing \rwrt\ we assume the
boundary $\p V_N$ is invariant under the action of $J^3_R$ so that
$h_p$ can be taken to be a $J^3_R$ eigenform with eigenvalue
$j_R$.  
Using $ \{ D , i_D\}=K$ together with \rwrt\ in \cqd\
yields \eqn\rfcx{E=||\p h_p||_{V_N}^2+||\p^\dagger h_p ||_{V_N}^2
+ ||\sqrt{K} h_p ||_{V_N}^2-2j_R||h_p ||_{V_N}^2-{\rm Re} \int_{\p
V_N}*D e^{-\phi}|h_p|^2.}

So far we have said nothing about the region $V_N$. If 
$V_N$ satisfies 
\foot{This condition on $V_N$
of course has a $j_R$ dependence, but we suppress this in the following
to avoid cluttering the equations.} 
\eqn\fty{ V_N \subset \{K > 2j_R+1\}, }
so that $V_N$ is near 
the region where one or more black holes are coincident,
then the sum of third and fourth terms in \rfcx\ will be greater than 
$||h_p||^2$. 
Also, note that the last term in \rfcx\ is nonnegative if the 
outward unit normal to $V_N$, call it $n^V$, obeys
\eqn\ftyy{ D^a n^V_a < 0 .}
So in this case
\eqn\bound{E(h_p)> ||h_p||_{V_N}^2 .}  
It is a theorem from complex analysis that the bound
\bound\ implies the vanishing of cohomology 
(see, e.g. section 4.4 of \krantz).
We therefore conclude that if $V_N$ obeys
\fty\ and \ftyy, then the Neumann cohomology of
$V_N$ is trivial, \eqn\hv{H^p(V_N)=0.}  
The simplest example of such a region is just 
$V_N = \{K > 2j_R+1\}$.

We now consider the opposite case, namely a region $W_N$ such that  
\eqn\ftx{W_N \subset \{ K < a\},} for some constant $a$.  In this case,
we may conjugate the differential operator $\p-D$ by 
$e^K$ to get $e^{-K} (\p-D) e^K = \p$, the usual differential
operator.  The function $e^K$ is bounded on $W_N$, so the cohomology
is unchanged.\foot{This deformation of the differential operator may be 
equivalently viewed as multiplication of the metric by some function of $K$,
which is bounded on $W_N$. 
This new metric is quasiisometric to the old, so the cohomology is unchanged.
}
We therefore need to show that the functional
\eqn\enfun{E=||\p h_p||_{W_N}^2 + ||\p^\dagger h_p||_{W_N}^2}
satisfies the appropriate bound.
First, note that for any number $\alpha$ we may follow the logic of 
\cqd\ to get 
\eqn\ddkk{\eqalign{
E= 
~ &||(\p -\alpha D) h_p||_{W_N}^2+||(\p^\dagger-\alpha i_D) h_p||_{W_N}^2 \cr
+&2\alpha {\rm Re} \langle h_p \{ \p, i_D\} | h_p \rangle_{W_N} 
- \alpha^2 \langle h_p \{ D, i_D\} | h_p \rangle_{W_N} \cr
}}
so that by \ftx\
\eqn\enbnd{E \ge 
2\alpha {\rm Re} \langle h_p \{ \p, i_D\} | h_p \rangle_{W_N} 
- \alpha^2 a ||h_p ||_{W_N}^2.}
Let us further demand that 
\eqn\ftxx{ D^a n^W_a > 0 ,}
where ${n^W}$ is the outward directed normal to $W_N$.
The integral 
$\int_{\p W_N}*D e^{-\phi}|h_p|^2$ is positive, so \rwrt\ implies that 
\eqn\lglg{\eqalign{ Re \langle &h_p\{\p,i_D\}| h_p\rangle 
\ge j_R ||h_p||^2_{W_N}
}}
for any $j_R$ and $h_p$ obeying Neumann boundary conditions.
Equation \enbnd\ then becomes
\eqn\enbndd{E \ge 
(2\alpha j_R - \alpha^2 a) ||h_p ||_{W_N}^2.}
For any $j_R >0$ we may choose an $\alpha$ small enough that 
$(2\alpha j_R - \alpha^2 a)>0$, so 
\enbndd\ implies the vanishing of cohomology.
We thus conclude that if the region $W_N$ satisfies \ftx\ and \ftxx,
then the positive charge Neumann cohomology of
$W_N$ is trivial, \eqn\hw{H^p(W_N)=0 ~~~~~~~~{\rm for \ } j_R > 0.}
The simplest example of such a region is just 
$W_N = \{K < a\}$
for some constant $a$.

\subsec{Exact Sequences Relating Subsets of $\m_N$}
We will now use these two vanishing theorems to study the cohomology
of $\m_N$.
Consider a $W_N$ satisfying \ftx\ and \ftxx.
Hodge duality 
sends $p\to n-p$, $(j_L,j_R) \to (-j_L,-j_R)$ and
interchanges Neumann and Dirichlet boundary conditions, 
so\foot{One must take care when applying Hodge duality to cohomology 
with respect to the operator $\p-D$.  
If a harmonic form $h$ is annihilated by $\p-D$ and
$\p^\dagger - i_D$ then $*h$ is annihilated by $\p+D$ and
$\p^\dagger + i_D$.  So in general Hodge duality will not interchange
cohomology classes.  However, on regions where $K$ is bounded then we may
multiply forms by a factor of $e^K$ and reduce to usual 
$\p$-cohomology.  Thus Hodge duality allows us to relate 
Dirichlet and Neumann cohomology on
$W_N$, but not on $V_N$ or $\m_N$.
}
\eqn\ser{H^p(W_N, j_L,j_R) = H^{n-p}_D(W_N,-j_L, -j_R).}  
The exact sequence of forms 
\eqn\aaa{
0\lra\Omega_D(W_N,j_L,j_R)\lra\Omega(\m_N,j_L,j_R)
\lra\Omega(\m_N\setminus W_N,j_L,j_R)\lra0
}
induces a long exact sequence relating the cohomology
of $W_N$ to that of its complement $\m_N\setminus W_N$.  
However, if we choose $W_N$ such that
$\m_N\setminus W_N$ satisfies \fty\ and \ftyy\ 
then our vanishing theorem on $\m_N\setminus W_N$ assures us that 
\eqn\kkdd{\eqalign{
H^p(\m_N,j_L,j_R) &= H^p_D(W_N,j_L,j_R) \cr
		  &= H^{n-p}(W_N,-j_L,-j_R) .
}}

For future reference, let us apply this formula in the two black hole case,
with $W_2=\{\x : |\x|>c\}$.  Plugging in the result \trg\ we find that
$W_2$ has nonzero cohomology
\eqn\wco{
H^1(W_2,j_L,j_R) = C,~~~~~~ {\rm for \ } j_R < 0, ~~|j_L| < |j_R|,
~~~ {\rm and} ~ j_L + j_R + \half \in {\bf Z}.
}

We will also make use of a slightly modified construction.
For a choice of $V_N$ and $W_N$ that satisfies \fty, \ftyy, \ftx\ and \ftxx, 
and whose union is the entire space $\m_N,$ define the region
$Y_N$ to be \eqn\yndef{Y_N = V_N \cap W_N.}
$Y_N$ resembles a shell surrounding the singularities of $K$.
We have the short exact sequence of complexes
\eqn\shor{0\lra\Omega(\m_N)\ra{r}\Omega(W_N)\oplus\Omega(V_N)
\ra{s}\Omega(Y_N)\lra0,}
where $r$ denotes the restriction map and $s$ denotes the subtraction map.
Both $r$ and $s$ are compatible with the differential operator $\p-D$, so
the usual arguments give the Mayer-Vietoris sequence for $\p-D$ cohomology,
\eqn\mvd{\cdots \lra H^{p}(\m_N)\lra H^{p}(V_N) \oplus H^{p}(W_N)\lra 
H^{p}(Y_N)\lra H^{p+1}(\m_N)\lra \cdots}
Plugging in \hv\ and \hw\ gives
\eqn\hm{H^{N-1}(\m_N)=H^{N-2}(Y_N) ~~~~~~~~{\rm for \ } j_R > 0.}
We saw in section 2.3 that all chiral primaries have $j_R>0$,
so \hm\ gives the complete cohomology of $\m_N$.

\subsec{Exact Sequence for $\m_3$ Cohomology}

Let us apply the result \hm\ of the last subsection to compute the
cohomology of $\m_3$.  Define 
$W_3^c=\{{\bf \x}: ~|\x^1| > c, ~|\x^2| > c, ~|\x^1-\x^2| > 4c \}$
and $V_3^c = \{{\bf \x}: ~|\x^1| < 2c \} \cup
\{{\bf \x}: ~|\x^2| < 2c \} \cup \{{\bf \x}: ~|\x^1-\x^2| < 8c \}$.
For any value of $j_R$ we may choose a value of $c$ such that
these regions satisfy $V_3^c \cup W_3^c =
\m_3$ as well as the conditions \fty, \ftyy, \ftx, \ftxx.   

We identify $Y_3$ as the union of three subspaces $U_1,U_2,U_3,$ and use 
Mayer-Vietoris sequences to compute its cohomology.  The three subspaces
are defined as
\eqn\thrys{\eqalign{
U_1 &= W_3^c \cap \{{\bf \x}: ~|\x^1| < 2c \} \cr
U_2 &= W_3^c \cap \{{\bf \x}: ~|\x^2| < 2c \} \cr
U_3 &= W_3^c \cap \{{\bf \x}: ~|\x^1-\x^2| < 8c \}.}}
Geometrically, the region $U_1~(\cong U_2)$ looks like a thickened cylinder 
with a single hole removed, 
whereas $U_3$ looks like a thickened cylinder with two holes removed. 
Notice that $U_1 \cap U_2 = \emptyset$.  Therefore $H^p(U_1 \cup U_2) = H^p(U_1) \oplus H^p(U_2)$. The long exact sequence derived from $Y_3 = (U_1 \cup U_2) \cup U_3$, which is
\eqn\smun{
\cdots \rightarrow H^p(Y_3) \rightarrow H^p(U_1 \cup U_2) \oplus H^p(U_3) \rightarrow H^p((U_1 \cup U_2) \cap U_3) \rightarrow H^{p+1}(Y_3) \rightarrow \cdots
}
reduces to
\eqn\biun{
 \cdots \rightarrow H^p(Y_3) \rightarrow H^p(U_1) \oplus H^p(U_2) \oplus
H^p(U_3) \rightarrow H^p(U_1 \cap U_3) \oplus H^p(U_2 \cap U_3) 
\rightarrow H^{p+1}(Y_3) \rightarrow \cdots
}
So we must compute the cohomology of the three spaces $U_1~(\cong U_2)$,
$U_3,$ and $U_1 \cap U_3~~(\cong U_2 \cap U_3)$.  
One more definition will help.  For $i=1,2$, let
\eqn\fridol{
U_3^i = \{{\bf \x}: ~4c < |\x^1-\x^2| < 8c, ~|\x^i| > c \}.
}
This includes a region that is singular with respect to the usual metric
on $\m_3$; for example, 
$U_3^1 = U_3 \cup 
\{{\bf \x}: ~4c < |\x^1-\x^2| < 8c, ~|\x^1| > c, ~|\x^2|\le c \}$
contains the $\x^2=0$ singularity.
We will remedy this by taking the metric on the $|\x^2|\le c$ component to 
be non-singular; we replace the $(d\x^2)^2\over |\x^2|^4$ term in the metric
by $(d\x^2)^2$.
Roughly speaking, $U_3^1$ is a cylindrical tube with a hole removed at $|\x^1| \leq c$.  With the change of variables $\x^1\to \x^1-\x^2$, 
$\x^2\to\x^1$ this region looks like $U_2$, and in fact with the
non-singular metric on $U_3^1$ these two regions are quasi-isometric.
We conclude that for the purposes of cohomology, 
$U_3^i \cong U_1 \cong U_2$.  
Moreover, note that
$U_3^1 \cap U_3^2=U_3$ with the correct $\m_3$ metric, so
\eqn\liun{
\cdots \rightarrow H^{p-1}(U_3) \rightarrow H^p(U_3^1 \cup U_3^2) 
\rightarrow H^p(U_3^1) \oplus H^p(U_3^2) \rightarrow H^p(U_3) 
\rightarrow \cdots
}

First, let us compute the $U_3^i$ cohomology.  $U_3^i$ is the
product of $B^*$ (a punctured four ball) 
and $\{{\bf \x}: ~|x^i| > c\}$.  This base is $W_2$,
so the K\"unneth formula gives
\eqn\sklo{\eqalign{
H^p(U_3^i) &= \bigoplus_{q} H^q(B^*) \otimes H^{p-q}(W_2) \cr
&\simeq H^0(B^*) \otimes H^p(W_2) \cr
}}
where we use the fact that $W_2$ has cohomology only at negative $j_R$,
so that only the positive charge cohomology of $B^*$ will contribute 
(see appendix B) if we evaluate $H^p(U_3^i)$ at nonnegative $j_R$.
We conclude from \wco\ that 
the only nonvanishing cohomology is at $p=1.$  But note further that 
we can replace $H^1(W_2)$ by $H^1(B^*)$.
Therefore the quantities $H^p(U_1) \oplus H^p(U_2)$ in \biun\ and 
$H^p(U_3^1) \oplus H^p(U_3^2)$ in \liun\ can both be replaced by 
$H^1(B^* \times B^*)$ for $p=1$ and $0$ for $p \neq 1$.

Second, we will use \liun\ to compute $H^p(U_3).$  
Let us see that $H^p(U_3^1 \cup U_3^2) = 0.$  
This space is a $B^*$ fibration over the space $S=\{{\bf \x}:~\x^1=\x^2\}$. 
We will show that $H^p(S)=0.$  For computing cohomology, 
the space $S$ can be identified with $B \cup W_2$.  
The intersection is $B \cup W_2 \simeq B^*$.  
From the preceding subsection, we know that cohomology of $W_2$ 
vanishes for non-negative $j_R$.  
Hence for charge $j_R\ge 0$ the relevant exact sequence reduces to 
\eqn\seofs{\eqalign{
0 \rightarrow H^0(S) \rightarrow H^0(B) \rightarrow H^0(B^*) \rightarrow
H^1(S) \rightarrow H^1(B) \rightarrow H^1(B^*) \cr \rightarrow H^2(S) 
\rightarrow H^2(B) \rightarrow H^2(B^*) \rightarrow 0.
}}
From the results of appendix B it follows that $H^2(S)=H^1(B^*)$, which
is zero for $j_R\ge 0$.  
Moreover, the map $H^0(B) \rightarrow H^0(B^*)$ is simply the restriction map.
Hartog's theorem states that in complex dimension greater than 
1, holomorphic functions on a domain $D$ minus an interior 
ball extend across the entire domain $D$, hence this restriction map is an
isomorphism.
This implies that $H^0(S)=H^1(S)=0$, so $H^p(U_3^1 \cup U_3^2) = 0$.
At negative $j_R$ we apply Hodge duality to the positive $j_R$ results, 
and conclude that $H^p(S)=0$ for all $j_R$.
Substituting the results of the previous two paragraphs into \liun, 
we find that the only nonvanishing cohomology of $U_3$ is 
\eqn\slue{
H^1(U_3) = H^1(B^* \times B^*).
}

Third, we notice that 
\eqn\homrhu{
H^p(U_1 \cap U_3) = H^p(B^* \times B^*).
}
This is clear because $U_1 \cap U_3$ is defined by the conditions
$c<|\x^1|<2c$ and $4c<|\x^1-\x^2|<8c$; these already imply the third condition, $|\x^2|>c$.  So $U_1 \cap U_3 \cong U_2 \cap U_3$ 
is biholomorphic to $B^* \times B^*$.

Now we are ready to substitute all these results into \biun.  Recall from the preceding section that the only nonvanishing cohomology of $Y_3$ is at $p=1.$  The long exact sequence reduces to
\eqn\subiun{
0 \rightarrow 2H^0(B^* \times B^*) \rightarrow H^1(Y_3) \rightarrow 2H^1(B^* \times B^*) \rightarrow 2H^1(B^* \times B^*) \rightarrow 0.
}
Therefore $H^1(Y_3) \simeq 2H^0(B^* \times B^*)$.  Using \hm, we conclude that 
\eqn\xhmg{
H^2(\m_3) \simeq 2H^0(B^* \times B^*).
}

It is useful to further refine the equation \xhmg. The
cohomology appearing in any exact sequence can be restricted to
eigenspaces of an operator which commutes with $\p-D$. In the
case at hand, for $p$-th cohomology, 
two such operators are $J^3_L$ and $J^3_R-{p \over 2}$, whose
eigenvalues are denoted $j_L$ and $j_R-{p \over 2}$.  
We then have the relation
\eqn\xhmjj{
H^2(\m_3,j_L,j_R) = 2H^0(B^* \times B^*,j_L,j_R-1).
}
In this expression the
second and third arguments of the cohomology groups
are the $J^3_L$ and $J^3_R$
eigenvalues, respectively.

\subsec{The Index} 

The partition function for three black holes is, according to \xhmjj, 
\eqn\pfn{{\cal Z}^{(3)}(y,z)=\sum
{\rm dim} H^{2}(\m_3,j_L, j_R)z^{2j_R}y^{2j_L}
=2 \left[{\cal Z}^{(2)}(y,z)\right]^2 
}
and the superconformal index is
\eqn\pfn{{\cal I}^{(3)}(y)=\sum {\rm dim}
H^{2}(\m_3,j_L, j_R)(-)^{2j_R}y^{2j_L}
= 2 \left[{\cal I}^{(2)}(y)\right]^2.
}

\newsec{Super-Poincare Cohomology}

In this paper we have defined the index ${\cal I}^{(N)}$ 
by exploiting the enhanced superconformal structure of 
low-energy black hole quantum mechanics. One may also define the index 
using only the superpoincare structure. This leads to the standard
formula for the Witten index in supersymmetric quantum mechanics 
as the dimension of the kernel of $\p+\p^\dagger$ minus the dimension of
the kernel of the adjoint. In order to make this well defined 
we must restrict to $L_2$ forms (without an $e^{-2K}$ measure factor). 

It is not hard to see, at least for $N=2$, that there are no $L_2$ 
eigenstates of $(\p+\p^\dagger)^2$ and this definition leads to 
a trivial index, 
in contrast to the superconformal result \indf. This comes about because  
the states which contribute to the index as computed in \indf\ in some
sense live at the boundary of moduli space 
(where black holes coincide) and are lost 
in the restriction to $L_2$ states. 

Potentially, these different definitions of the index are answers to 
different physical questions. 
Which definitions of the index will be useful for 
a full 
understanding of low-energy black hole dynamics remains to be
seen.

\centerline{\bf Acknowledgements}

We are grateful to J. Maldacena, J. Michelson, M. Spradlin, 
C. Vafa, A. Volovich and M. Wijnholt
for useful conversations. This work is
supported in part by DOE grant DE-FG02-91ER40654, NSF grant PHY99-07949, 
and an NSF graduate fellowship.  M. S. is supported by NSF grant 
DMS 9870161.

\appendix{A}{The Behavior of $L$}

In this appendix we detail some properties of $L$ and verify the
bound \eest\ used in the proof of the first vanishing theorem.

The expression \gtf\ for the function $L$ is divergent, but this
divergence disappears after differentiating $L$ to form the
metric. The irrelevant divergences can be subtracted, yielding the
finite (for $\vec X^A \neq \vec X^B$), but more
complicated-looking, expression \jmas\ \eqn\ltg{L=L_2+L_3,} with
\eqn\frds{\eqalign{L_2&=-6\pi^2Q^3\sum_{A\neq B}^N{\ln |\vec X^A-\vec
X^B|\over  |\vec X^A-\vec X^B|^2},\cr L_3&=-Q^3\sum_{A\neq B \neq
C}^N\int d^4X {1 \over  |\vec X-\vec X^A|^2 |\vec X-\vec X^B|^2
|\vec X-\vec X^C|^2}.}} These obey ${\cal L}_D L_3 =L_3$ and 
${\cal L}_D L_2 =L_2+\half K$.   

We wish to show that the quantity
\eqn\frtz{{|\p^- L|^2 \over K \ln^2 K}} is bounded in the regions
of small $K < {1 \over \mu}$ and large $K>\mu$ for 
sufficiently large $\mu$.  

On a surface of constant $K$
everything is bounded.  
Let us denote by $\Theta$ a set of coordinates on such a surface.
The Lie derivative $\cl_D$ generates 
motion to larger values of $K$ as 
${\cal L}_D = K \ d/dK$.
First, note that $\cl_D (L/K) = \half$, which may be integrated 
along orbits of $\cl_D$ to give 
\eqn\Lis{{L\over K} = \half \ln K + f(\Theta)}
where $f$ is some function of the angular coordinates. 
Now, using $I_-=-\p\p^-L$ we find that 
\eqn\ldlm{\cl_D { |\p^-L|^2 \over K} = {L \over K} + \half.}
Plugging in \Lis\ we integrate to find 
$|\p^-L|^2/K = {1\over 4} \ln^2K + f(\Theta) \ln K + g(\Theta)$
where $f$ and $g$ are both functions only of $\Theta$.
We conclude that 
\eqn\boundl{
{|\p^- L|^2 \over K \ln^2 K} = 
{1\over 4} + {f(\Theta) \over \ln K} + {g(\Theta)\over \ln^2K}.
}
This goes to $1\over 4$ at $K\to 0$ and at $K\to \infty$
since $f$ and $g$ are both bounded functions.  Thus \frtz\ 
is bounded at both small $K$ and large $K$. 

\appendix{B}{Ball Cohomology}

We now summarize various results on ball cohomology in four dimensions.  
As usual we break the
cohomology groups into eigenspaces of $J^3_L$ and $J^3_R-{p \over 2}$.
Recall that we use $\p$-cohomology rather than $\bp$-cohomology.

Let $B$ be the standard 4-ball.  Then the only nonzero Neumann cohomology is
\eqn\aax{
H^0(B,j_L,j_R) = {\bf C}, ~~~~~~~~~~ j_R \ge 0, ~~~ |j_L| \le |j_R|, 
~~~ {\rm and} ~ j_L+j_R\in {\bf Z}.
}
This may be seen by noting that anti-holomorphic functions on the 4-ball 
are generated by monomials of the form $(\bz_1)^a (\bz_2)^b$, where 
$j_R=\half(a+b)$ and $j_L = \half(a-b)$.
Here $(z_1,z_2)$ are the usual complex coordinates on $R^4$.
Hodge duality exchanges Neumann and Dirichlet boundary conditions and
takes $(j_L,j_R) \to (-j_L,-j_R)$.
Thus the nonvanishing Dirichlet cohomology of $B$ is 
\eqn\aay{
H_D^2(B,j_L,j_R) = {\bf C}, ~~~~~~~~~~ j_R \le 0, ~~~ |j_L| \le |j_R|,
~~~ {\rm and} ~ j_L+j_R\in {\bf Z}.
}

Let $B^*$ be the punctured ball, which is the standard
4-ball minus a smaller 4-ball centered at the origin.  
We use the short exact sequence relative to this decomposition
\eqn\aaa{
0\lra\Omega_D(B,j_L,j_R)\lra\Omega(B,j_L,j_R)\lra\Omega(B^*,j_L,j_R)\lra0
}
which relates forms with Dirichlet and Neumann boundary conditions.
As usual, this induces a long exact sequence giving the Neumann cohomology
of $B^*$ in terms of \aax\ and \aay.  The result is 
\eqn\pbs{\eqalign{
H^0(B^*,j_L,j_R) &= {\bf C}, ~~~~~~~~~~ j_R \ge 0, ~~~ |j_L| \le |j_R|,
~~~ {\rm and} ~ j_L+j_R\in {\bf Z} \cr
H^1(B^*,j_L,j_R) &= {\bf C}, ~~~~~~~~~~ j_R \le -\half, ~~~ |j_L| \le |j_R|,
~~~ {\rm and} ~ j_L+j_R\in {\bf Z}.
}}

\listrefs

\end